# A New Macho Search Strategy

**Andrew Gould**[*]

Dept of Astronomy, Ohio State University, Columbus, OH 43210

e-mail gould@payne.mps.ohio-state.edu

## Abstract

I propose a radical revision in the search strategy for Massive Compact Objects (Machos) toward the Galactic bulge: monitor the entire $\sim 300$ square deg of the bulge and tune the search primarily to microlensing events of bright $M_V \lesssim 2$ stars. By employing a small second telescope to follow up the events detected with low dimensionless impact parameter $\beta \lesssim 0.2$, one could measure the proper motion $\Omega$ of $\sim 10\,\mathrm{yr}^{-1}$ events and significantly constrain $\Omega$ for a similar number. If, in addition, the events were followed from a Macho parallax satellite, it would be possible to measure individual masses, distances, and transverse velocities for the events with proper motions. If a fraction $\xi$ of the bulge is composed of low-mass objects in the range $10^{-3} \lesssim M_{\mathrm{low}}/M_\odot \lesssim 10^{-1}$, then mass measurements could be made for $\sim 10(\xi/0.1)(0.01 M_\odot \langle M_{\mathrm{low}}^{-1} \rangle)\,\mathrm{yr}^{-1}$ of these, thus allowing a direct measurement of the mass function in the sub-stellar range. In the absence of a parallax satellite, ground-based observations could significantly constrain (but not measure) Macho parallaxes. These constraints, when combined with the proper-motion measurements, would in turn constrain the mass, distance, and transverse speed of the Machos. The proposed strategy should therefore be adopted even before a parallax satellite is launched.

Subject Headings: astrometry – gravitational lensing

## 1. Introduction

Massive Compact Objects (Machos) are currently being detected toward the Galactic bulge at a rapid rate by two groups. To date the MACHO collaboration has detected $\sim 28$ candidate events (Alcock et al. 1994, D. Bennet 1994, private communication) and the OGLE collaboration (Udalski et al. 1994) has detected $\sim 10$. This event rate, which is far higher than was anticipated, implies an optical depth of $\tau \sim 3\text{--}4 \times 10^{-6}$. At present, it is not known whether Machos are primarily in the disk, the bulge, or some other structure such as a highly flattened halo. Even the observation of several hundred events over many bulge fields would give only crude information on the mass distribution (Han & Gould 1994b). Addressing more subtle questions such as determining the mass spectrum and the velocity distribution would be essentially impossible using present techniques.

By detecting a Macho from its light curve, one determines the time scale, $\omega^{-1}$ which is a combination of the Macho mass $M$, distance $D_{\rm OL}$, and relative transverse speed, $v$:

$$\omega^{-1} = \frac{r_e}{v}, \qquad r_e^2 \equiv \frac{4GM D_{\rm OL} D_{\rm LS}}{c^2 D_{\rm OS}}. \tag{1.1}$$

Here $r_e$ is the Einstein radius, and $D_{\rm OL}$, $D_{\rm LS}$, and $D_{\rm OS}$ are the distances between the observer, lens, and source. There is a fourth parameter that one would also like to know, the angle of the transverse motion, $\Phi$. It does not enter the time scale. The distance to the source $D_{\rm OS}$ is regarded as known, either from the fact that it lies in a population of known distance, e.g. the Large Magellanic Cloud (LMC), or from spectroscopic parallax.

Two other types of measurements can yield three other parameters. Macho parallaxes give the "reduced" Einstein radius, $\tilde{r}_e$,

$$\tilde{r}_e \equiv \frac{D_{\rm OS}}{D_{\rm LS}} r_e. \tag{1.2}$$

(Gould 1992, 1994b, 1994c). Together with the time scale, this implies a measurement of the "reduced" transverse speed, $\tilde{v} = \omega \tilde{r}_e = (D_{\rm OS}/D_{\rm LS})v$. In some cases, parallax measurements also give the direction $\Phi$. While under exceptionally favorable circumstances it is possible to measure $\tilde{r}_e$ from the ground (Gould, Miralda-Escudé, & Bahcall 1994), generally a satellite telescope in solar orbit is required.

The proper motion $\Omega = v/D_{\rm OL}$ can be determined by measuring $\theta_*$ the angular size of the Einstein ring,

$$\theta_* = \frac{r_e}{D_{\rm OL}}, \qquad (1.3)$$

since $\Omega = \omega\theta_*$. To date, five methods have been advanced for measuring $\theta_*$: 1) resolving the two images (Gould 1992), 2) measuring the apparent motion of the centroid of the two images (Hog, Novikov, & Polnarev 1994), 3) deviations of the light curve from the standard form when the Macho transits the face of the star (Gould 1994; Nemiroff & Wickramasinghe 1994), 4) Color variation of the light curve during transits or near-transits due to limb darkening (Witt 1994; D. Welch 1994, private communication), and 5) apparent changes in the radial velocity during the event (Maoz & Gould 1994). Methods (1) and (2) give $\Phi$ as well as $\theta_*$, the rest give only $\theta_*$. Of these methods, (1) is applicable only to very massive Machos $M \gtrsim 300\,M_\odot$ even with the resolution of the *Hubble Space Telescope*. Method (2) requires hundreds of observations each with an accuracy $\sim \theta_*$, so that even with a dedicated 1m satellite telescope and bright sources, the mass must be $M \gtrsim M_\odot$. Method (5) can be applied even if the impact parameter is many stellar radii. However the source must be rotating fast enough that the line shifts due to differential magnification are noticeable. Moreover, the source rotation must be measurable which generally means that it must be faster than the turbulent velocity. These conditions restrict the sources to early type stars. The method is therefore well suited to observations toward the LMC (Alcock et al. 1993; Aubourg et al. 1993) where many of the sources are A stars, but it is not well suited to bulge observations (Han & Gould 1994a). Since Machos seen toward the bulge are of order $\sim 0.1\,M_\odot$, only methods (3) and (4) are applicable. Method (3) can be applied only if the Macho actually transits the face of the star (Gould 1994a). The range of method (4) has not been examined in detail, but I find that for bright sources, a measurement could be made with a small < 1m telescope for impact parameters within $\eta \sim 2$ stellar radii. That is,

$$\beta < \beta_{\max} = \eta \frac{r_s}{\theta_* D_{\rm OS}}, \qquad (1.4)$$

where $r_s$ is the source radius, $\eta$ is the sensitivity factor of the measurement, and $\beta$ is the impact parameter in units of the Einstein radius. Note that the fraction of events for which proper-motion measurements are possible is proportional to the size of the source.

## 2. Optimal Search Strategy

From a purely mathematical standpoint, an optimal search strategy would maximize $\sum_i r_{s,i}$, the sum of the physical radii of the source stars in detected events. This would maximize the number of events for which proper-motion measurements could be made. However, two practical considerations lead to a somewhat different strategy. First, for lensing events with small $r_s$, proper motions can be measured only for the very small fraction with $\beta \ll 1$. For example, for $r_s \sim 2r_\odot$ and typical parameters, one finds $\beta_{\max} \sim 0.02$. It is much more difficult to predict from the early light curve which events will have such extremely low impact parameters than it is to predict which will have $\beta \lesssim 0.1$ appropriate for stars with $r_s \sim 10 r_\odot$. Hence, in order to measure a few proper motions, one would have to invest a great deal of telescope time monitoring many events that contain little information. Second, the main value of a proper-motion measurement can be realized only if there is also a parallax measurement. If parallax photometry measurements are made at a rate $f = N\omega$ with accuracy $\sigma$, then the determination of $\tilde{r}_e$ will be degenerate unless $(N/10)^{-1/2}(\sigma/0.01) \lesssim \omega^{-1}/10\,{\rm days}$ (Gould 1994c). For big stars (which are for the most part also bright) this accuracy is not difficult to achieve. However, for small stars such accuracy would tax the capacities of a small satellite, particularly if it had to follow many events for each one with a usable proper motion.

Hence, the optimal strategy is to monitor the most stars with large radii. Using the light distribution as measured by Dwek et al. (1994) to scale from the luminosity function in Baade's Window as measured by Terndrup (1988), I estimate that the entire Galactic bulge contains $\sim 4 \times 10^6$ stars with $M_V < 2$. The best strategy is simply to monitor all of them. Since the bulge covers an area of $\sim 300$ square deg and since the largest camera used in current microlensing searches has a field of only 0.5 square deg, it is obvious that such a strategy would require significant modifications of both equipment and protocol. For example, if the focal length of the MACHO telescope were cut in half, it would cover 2 square deg with $\sim 1\rlap.{''}2$ pixels. These would be adequate to photometer bulge giants. The bulge could then be covered in 150 frames. If necessary, the exposure time in most fields could be substantially reduced from the current 150 seconds.

Even fields that are relatively heavily reddened should be observed. Present strategies avoid obscured areas of the bulge in order to maximize the total number of stars covered for a given amount of observing time. However, the bright bulge stars $M_I \lesssim 1$ that are the targets of this search strategy will have apparent magnitudes $I \lesssim 18$ even in fields with $A_V \sim 4$. They will be isolated in the sense that the surface density of comparable or brighter stars is $\sim (10'')^{-2}$. Hence, the photometry will be photon and flat-fielding limited (rather than crowding limited) so that it should be possible to get $\sim 1\%$ accuracy on an $I = 18$ star in $\lesssim 80\,\mathrm{s}$. Most fields would require even shorter exposures. In addition, the heavily reddened fields currently being avoided could be used to probe the inner part of the Galactic disk, a region which is not well sampled under the present strategy.

## 3. Proper Motion Measurements

If $\mathcal{N} = 4 \times 10^6$ stars were monitored over a bulge season of duration $T = 180$ days, then the number of lensing events with impact parameters $\beta < 1$ Einstein radius would be $(2/\pi) \langle \omega \rangle \tau \mathcal{N} T$ where $\langle \omega \rangle$ is the mean inverse time scale and $\tau$ is the optical depth. For Machos characterized by a distance $D \equiv D_{\rm OS} D_{\rm LS} / D_{\rm OL}$ and mass $M$, the number of proper-motion detections would therefore be [eq. (1.4)],

$$\begin{aligned} N &= \frac{2}{\pi} \eta (4GMD)^{-1/2} c \langle r_s \rangle \langle \omega \rangle \tau \mathcal{N} T \\ &= 5 \frac{\eta}{2} \left( \frac{M}{0.1 M_\odot} \right)^{-1/2} \left( \frac{D}{1\,\rm kpc} \right)^{-1/2} \frac{\langle r_s \rangle}{10 r_\odot} (10\,{\rm days}\,\langle \omega \rangle) \frac{\tau}{10^{-6}}. \end{aligned} \quad (3.1)$$

Thus, for bulge Machos, assuming $\tau \sim 2 \times 10^{-6}$ and $D_{\rm LS} \sim 1\,{\rm kpc}$, there would be $N \sim 9\,{\rm yr}^{-1}$ proper-motion measurements. For disk Machos assuming $\tau \sim 1 \times 10^{-6}$ and $D_{\rm LS} \sim 4\,{\rm kpc}$, there would be $N \sim 2\,{\rm yr}^{-1}$ measurements.

In making the above estimates, I have taken the typical Macho mass to be $M \sim 0.1\,M_\odot$. If there is a sub-population of Machos which have substantially lower characteristic mass $M_{\rm low}$ and which comprise a fraction $\xi$ of the total bulge mass, these will give rise to a fraction $\sim \xi (M_{\rm low}/M_\odot)^{-1/2}$ of all events, and hence a fraction $\sim \xi (M_{\rm low}/M_\odot)^{-1}$ of all proper-motion measurements. Thus, if there is a dynamically significant low-mass population, its mass spectrum can be directly measured.

## 4. Physical and Observable Parameters

The relations between the observable parameters ($\omega$, $\tilde{r}_e$, $\theta_*$, and $D_{\rm OS}$) and the physical parameters ($M$, $D_{\rm OL}$ and $v$) are as follows:

$$M = \frac{c^2}{4G}\tilde{r}_e\theta_*, \tag{4.1}$$

$$\frac{D_{\rm OL}}{D_{\rm OS}} = \left(\frac{\theta_* D_{\rm OS}}{\tilde{r}_e} + 1\right)^{-1}, \tag{4.2}$$

and

$$v = \frac{\omega}{\tilde{r}_e^{-1} + (\theta_* D_{\rm OS})^{-1}}. \tag{4.3}$$

## 5. Importance on Non-Detections

Equations (4.1)–(4.3) can be used to illustrate the value of non-detections as well as measurements of Macho proper motions, assuming that a parallax measurement has been made. For example, suppose that the actual parameters of a Macho are $D_{\rm OL} = 2\,{\rm kpc}$, $M = 0.1\,M_\odot$, $v = 100\,{\rm km\,s^{-1}}$, and that the source star lies at $D_{\rm OS} = 8\,{\rm kpc}$ and has a radius $r_s = 10\,r_\odot$. The event is measured as having time scale $\omega^{-1} = 19.0\,{\rm days}$ and $\beta = 0.04$. The reduced Einstein radius is measured to be $\tilde{r}_e = 2.2 \times 10^{13}\,{\rm cm}$, corresponding to $\tilde{v} = 133\,{\rm km\,s^{-1}}$. An attempted proper-motion measurement with sensitivity $\eta = 2$ is made, but fails. From equation (1.4), one then finds that $\theta_* D_{\rm OS} > 3.5 \times 10^{13}\,{\rm cm}$. Hence, $M > 0.053\,M_\odot$, $D_{\rm OL} < 3.1\,{\rm kpc}$, and $v > 82\,{\rm km\,s^{-1}}$. In addition, $v < \tilde{v} = 133\,{\rm km\,s^{-1}}$. One would therefore learn that Macho was definitely not in the bulge and that its speed was of order $100\,{\rm km\,s^{-1}}$, i.e., consistent with disk kinematics. Or consider another example of a bulge Macho with $D_{\rm LS} = 0.8\,{\rm kpc}$, $M = 0.1\,M_\odot$, $v = 200\,{\rm km\,s^{-1}}$, and a source star again at at $D_{\rm OL} = 8\,{\rm kpc}$ and $r_s = 10\,r_\odot$. The observables are $\omega^{-1} = 6.59\,{\rm days}$, $\beta = 0.166$, $\tilde{r}_e = 1.14 \times 10^{14}\,{\rm cm}$, $\tilde{v} = 2000\,{\rm km\,s^{-1}}$. No proper motion is detected with $\eta = 2$.

In this case, one can place a limit $v \lesssim 600\,\mathrm{km\,s^{-1}}$ just by assuming that the Macho is bound to the galaxy. Hence, from equation (4.3), $\theta_* D_{\mathrm{OS}} \lesssim 4.9 \times 10^{13}\,\mathrm{cm}$, so that $D_{\mathrm{LS}} \lesssim 2.4\,\mathrm{kpc}$ and $M \lesssim 0.39\,M_\odot$. On the other hand, from the non-detection of a proper motion $D_{\mathrm{OS}}\theta_* > 8.4 \times 10^{12}\,\mathrm{cm}$, implying $D_{\mathrm{LS}} > 0.55\,\mathrm{kpc}$ and $M > 0.067\,M_\odot$. These examples illustrate that non-detections can place valuable constraints on the parameters of individual Machos.

## 6. Ground-Based Parallax Limits

For special classes of events which are either very long ($\omega^{-1} \gtrsim \mathrm{yr}/2\pi$) or for which the reduced speed is very low ($\tilde{v}/v_\oplus \lesssim 3$) it is possible to measure the reduced velocity from the ground (Gould 1992; Gould et al. 1994). Unfortunately, these events are so rare and also so atypical that they provide little information about the Macho distribution as a whole. However, there is a much wider class of events for which it is possible to measure $\tilde{r}_+ = \tilde{r}_e \sec\phi$ where $\phi$ is an (unknown) angle to be described below. I refer to $\tilde{r}_+$ as the "parallax limit", since $\tilde{r}_e \leq \tilde{r}_+$. The observations required to measure $\tilde{r}_+$ are quite similar to the follow-up photometry needed to measure the proper motion. Moreover, the measurement is most feasible for low impact-parameter events, just the class that would be followed for proper motions. Hence, a large fraction of events with proper motions would also have parallax limits. What can be learned from simultaneous measurements of $\theta_*$ and $\tilde{r}_+$? From equations (4.1)–(4.3), these two quantities provide upper limits on $M$, $D_{\mathrm{OL}}$, and $v$. The uncertainty in $M$ is directly proportional to the uncertainty in $\cos\phi$ which, of course, varies from 0 to 1. For bulge Machos (the great majority of Machos for which proper motions can be obtained) the uncertainties in $D_{\mathrm{OL}}$ and $v$ also scale with $\cos\phi$. Hence, in one sense, parallax limits appear to provide little information. However, within the context of likelihood fits to the data, parallax limits do provide information because within any given the model, the prior probability of $\theta$ is known. In fact, for most models $\theta$ will be roughly uniformly distributed so that, for example, a large fraction of the mass upper limits will be near the true masses.

Parallax limits can be measured from the ground to the extent that the Earth's acceleration induces a noticeable change in the light curve. For $\tilde{v} \gg v_\oplus = 30\,\mathrm{km\,s^{-1}}$ and $\omega^{-1} \ll \mathrm{yr}/2\pi$ the fractional change in relative velocity (or time scale) as a function of time during the event is $\Delta\omega(t)/\omega = a_\parallel(t-t_0)/\tilde{v}$, where $t_0$ is the time of maximum magnification, $a_\parallel \equiv \mathbf{a}\cdot\hat{\mathbf{v}}$, and $\mathbf{a}$ is the two-dimensional projection of the Earth's acceleration vector onto the plane of the sky. For observations toward the bulge, $\mathbf{a}$ is usually aligned closely with the ecliptic. The angle $\phi$ is defined by $\cos\phi = a_\parallel/a$. If $\Delta\omega/\omega$ can be detected, the quantity that is measured is therefore $\tilde{v}\sec\phi$ or equivalently $\tilde{r}_+ \equiv \tilde{v}\sec\phi/\omega$. Note that $\phi$ is related to $\Phi$ by a known offset between the fiducial $x$-axis and the direction of $\mathbf{a}$.

I reserve the discussion of the conditions under which parallax limits can be measured for a future paper. Here I simply remark that for bright stars and events discovered in the spring or fall with low impact parameters, the prospects for measuring $\tilde{r}_+$ are generally good.

## 7. Conclusions

In order to obtain the maximum amount of information about Machos being detected toward the Galactic bulge, it is necessary to measure additional parameters. Individual masses and distances can be determined only by measuring both the parallax and the proper motion of the Macho. While parallaxes can be measured for a large fraction of the events from a satellite, the only realistic hope for measuring significant numbers of proper motions is by precision photometry of events where the Macho passes over or near the face of the star. The Macho search strategy should therefore be reoriented toward observing the maximum number of stars with large radii. That is, the entire $\sim 300$ square deg of the bulge should be monitored with observations tailored to find the lensing events of stars $M_V < 2$. About 10 proper motions could then be measured per year, more if there is a substantial population of low-mass objects $M \ll 0.1 M_\odot$. This strategy is advocated for the bulge only. It should not be applied to searches toward the LMC.

**Acknowledgements**: I would like to thank D. DePoy, C. Han, D. Terndrup, and D. Welch for valuable discussions.